\documentclass[twoside]{aiml14}

\usepackage{aiml14macro}

\usepackage{amsmath,xspace}
\usepackage{stmaryrd}
\usepackage{subfigure}
\usepackage{relsize}
\usepackage{hyperref}

\usepackage{tikz}
\usetikzlibrary{arrows,shapes,shapes.arrows,shapes.callouts,decorations.pathmorphing,backgrounds,positioning,fit,calc}
\tikzset{
  every picture/.style={inner frame sep=0pt
                       }
 ,n/.style={circle,fill,draw,inner sep=1.5pt,node distance=1.5cm}
 ,root/.style={circle,draw,inner sep=1.5pt,node distance=1.5cm}
 ,nbhd/.style={dotted,rounded corners=5pt,draw,node distance=1.5cm}
 ,arr/.style={->, >=stealth, semithick, shorten <= 3pt, shorten >= 3pt}
 ,arrnb/.style={arr,dotted}
 }

\renewcommand{\theta}{\vartheta}

\newcommand{\ATLdiamond}[1]{{\langle\!\langle#1\rangle\!\rangle}}
\newcommand{\Lang}{\mathcal{L}}

\newcommand{\NN}{\mathbb{N}}

\newcommand{\Land}{\bigwedge}
\newcommand{\Lor}{\bigvee}
\newcommand{\hearts}{\heartsuit}
\newcommand{\M}{\hearts}

\newcommand{\Rules}{\mathcal{R}}

\newcommand{\Var}{V}

\newcommand{\EL}{\ensuremath{\mathcal{EL}}\xspace}
\newcommand{\FL}{\ensuremath{\mathcal{FL}_0}\xspace}

\newcommand{\CL}{\ensuremath{\mathrm{CL}}\xspace}

\newcommand{\KD}{\ensuremath{\mathit{KD}}\xspace}
\newcommand{\MDN}{\ensuremath{\mathit{M_s}}\xspace}

\newcommand{\cset}[1]{{\left\{ #1 \right\}}}
\newcommand{\tup}[1]{{\langle #1 \rangle}}
\newcommand{\bbrack}[1]{{\llbracket #1 \rrbracket}}
\newcommand{\Sem}[1]{\bbrack{#1}}
\newcommand{\+}[1]{\mathcal{#1}}

\newcommand{\Pow}{\mathcal{P}}
\newcommand{\fP}{\Pow} 
\newcommand{\fQ}{\mathcal{Q}} 
\newcommand{\fB}{\+B_{\infty}} 
\newcommand{\fmN}{\+M}   
\newcommand{\Set}{\mathsf{Set}}

\newcommand{\FB}{\mathfrak{B}}
\newcommand{\FN}{\mathfrak{N}}

\newcommand{\natto}{\dot\to}
\newcommand{\into}{\hookrightarrow}

\newcommand{\Pos}{\mathsf{Pos}}
\newcommand{\Conj}{\mathsf{Conj}}
\newcommand{\ossub}{\sqsubseteq_1}

\makeatletter
\newcommand*{\@old@slash}{}\let\@old@slash\slash
\def\slash{\relax\ifmmode\delimiter"502F30E\mathopen{}\else\@old@slash\fi}
\makeatother

\def\lastname{Gor\'in and Schr\"oder}

\begin{document}

\begin{frontmatter}

\title{Subsumption Checking in{\\} Conjunctive Coalgebraic Fixpoint Logics}
\author{Daniel Gor{\'i}n}
\author{Lutz Schr{\"o}der}

\address{Friedrich-Alexander-Universit\"at Erlangen-N\"urnberg}

\begin{abstract}
  While reasoning in a logic extending a complete Boolean basis is
  coNP-hard, restricting to conjunctive fragments of modal languages
  sometimes allows for tractable reasoning even in the presence of
  greatest fixpoints. One such example is the \EL family of
  description logics; here, efficient reasoning is based on
  satisfaction checking in suitable small models that characterize
  formulas in terms of simulations. It is well-known, though, that not
  every conjunctive modal language has a tractable reasoning
  problem. Natural questions are then how common such tractable
  fragments are and how to identify them. In this work we provide
  sufficient conditions for tractability in a general way by
  considering unlabeled tableau rules for a given modal logic.  We work
  in the framework of coalgebraic modal logics as unifying semantic
  setting. Apart from recovering known results for
  description logics such as \EL and \FL, we
  obtain new ones for conjunctive fragments of relational and
  non-relational modal logics with greatest fixpoints.  Most notably
  we find tractable fragments of game logic and the
  alternating-time $\mu$-calculus.
\end{abstract}

 \end{frontmatter}

\section{Introduction}\label{sec:intro}

\noindent
The complexity of reasoning in logics extending a complete Boolean
basis is at least \textsc{coNP}. For modal logics, it is typically
even harder: already the basic (multi-)modal logic $K_m$ is
PSPACE-complete~\cite{Ladner77} and if fixed points are added to the
mix, the complexity typically goes up to at least EXPTIME which, e.g.,
is the complexity of PDL and the $\mu$-calculus~\cite{EmersonJutla99}.
Practical reasoning in these logics requires highly optimized
heuristic strategies and will ultimately have only a limited degree of
scalability.

This motivated the study of fragments in which core
reasoning problems become tractable, i.e.,
decidable in polynomial time. Such fragments typically exclude
negation and disjunction. Perhaps the best-known example is
the \EL family of \emph{lightweight description logics}, where
also universal restrictions (i.e.\ $\Box$-modalities) are excluded.
In the absence of negation, satisfiability is no longer the central
reasoning problem, being in fact often trivial (e.g., in \EL every
formula is satisfiable).  Instead, one focuses on the entailment
problem, alternatively called \emph{subsumption checking} in the DL
community.  Indeed \EL turns out to have a polynomial-time subsumption
problem~\cite{BaaderIJCAI03a,BaaderBrandtLutz05}, even when extended
with greatest fixed points~\cite{LutzEA10}. Despite the limited
syntax, \EL can in practice accommodate large ontologies such as
\textsc{Snomed~ct}.

Rather surprisingly, the subsumption problem of \FL,
the counterpart of \EL with universal instead of existential restriction,
becomes intractable when greatest fixed points (or even just non-recursive
global definitions, i.e.\ acyclic TBoxes) are added to the
language~\cite{Baader-AMAI-96,BaaderKuesters99,Nebel90}. This shows
that there is more to lightweight logics than just dropping
disjunctions.  Here, we aim to develop conceptual tools to identify
lightweight modal formalisms beyond the purely relational realm. For
uniformity, we work in the setting of \emph{coalgebraic modal
  logic}~\cite{Pattinson03}, where the notions of model and modal
operators are suitably abstracted. We then state and prove a general
version of each result and just obtain the featured instances as
corollaries.

Tractability of relational lightweight logics exploits the existence
of what are called \emph{materializations} of formulas (which moreover
need to be computable and small). (Alternatively, tractability can be
shown by proof-theoretic methods~\cite{Hofmann05}.) A materialization
for $\phi$ is a model that satisfies \emph{only} the formulas that
$\phi$ entails; thus, subsumption can be reduced to model-checking in
a materialization~\cite{LutzEA10}.  Moreover, while there seems to be
a strong connection between tractability of subsumption for a given
fragment and \emph{convexity} of its formulas, meaning that they imply
at least one of the disjuncts of every disjunction they entail, the
precise nature of such a connection is still partly unclear (see,
e.g.,~\cite{KrisnadhiLutz07,LutzWolter12}).

For coalgebraic logics, we show that a stronger (infinitary) version
of convexity of their conjunctive fragments is actually equivalent to
the existence of materializations (for the relational description
logic $\mathcal{ALCFI}$, a more fine-grained connection at the level
of TBoxes has been established by Lutz and
Wolter~\cite{LutzWolter12}).  As in the relational case, our
materializations moreover have an even stronger property --- they can
be taken as complete replacements of the materialized formulas, in the
sense that the satisfaction relation for the former corresponds to a
similarity relation for the latter in a sense we developed
recently~\cite{GorinSchroder13}.  However, the mere existence of
materializations is not enough for tractability; one requires
additional conditions that ensure that materializations for a given
conjunctive fragment of a modal logic $\Lang$ can be obtained in
polynomial time. For this, we develop a simple syntactic criterion
based on the set of (unlabeled) tableau rules for $\Lang$, and show
how to compute small materializations from them even in the presence
of greatest fixpoints. With this result we can show tractability of
several conjunctive modal (fixpoint) logics, including fragments of
game logic~\cite{Parikh83} and the alternating-time
$\mu$-calculus~\cite{AlurEA02}.

Proofs and more technical details can be found in the appendix.

\section{Preliminaries}\label{sec:prelim}

\noindent
We first present the various concrete logics that will serve as case
studies throughout the paper and then briefly introduce the basic
concepts of coalgebraic logic that are used in the generic
development. For each concrete logic we also consider its reasoning
principles, in the form of unlabeled tableau-rules
$\Gamma_0/\Gamma_1\mid \dots \mid\Gamma_n$, where the $\Gamma_i$ are
sets of formulas. A set of rules $\+R$ is meant to be used in the
usual way: to show satisfiability of a set $\Gamma$ (interpreted
conjunctively), one needs to show the satisfiability of at least
one conclusion of every rule in $\+R$ applicable to $\Gamma$. All
tableau systems are understood to extend a set of propositional rules.

\paragraph{Basic modal logic.} We assume the reader to be familiar
with the syntax and semantics of the basic modal logic $K$ interpreted
over Kripke models.  We shall also consider its restriction \KD to
\emph{serial} models, where every node has at least one successor, making $\Diamond \top$  valid.
The set of rules $\+R_K=\cset{K_n : n \geq 0}$ (Fig.~\ref{fig:tab}) induces a
complete tableau system for $K$. For \KD one needs to add to $\+R_K$
the rules $D_n$ for $n \geq 0$.

\paragraph{Monotone neighbourhood logic.} The minimal monotone logic
$M$ uses the same language as $K$ but is interpreted over monotone
neighbourhoods, i.e., neighbourhood models where the set of
neighbourhoods of each point is upwards closed w.r.t.\ set
inclusion~\cite{Chellas80}.  We read $\Box\phi$ as `there is a
neighbourhood where $\phi$ holds'.  It is well known that this logic
can be encoded in $K$, replacing $\Box$ with $\Diamond\Box$
and $\Diamond$ with $\Box\Diamond$ (e.g.~\cite{Parikh83}).  A
complete tableau system for $M$ is obtained simply by taking rule $K_1$.

Here too, we will be interested in the \emph{serial} case which
corresponds to the case where $\Box\top$ and $\Diamond\top$ are taken
as axioms; we shall denote the resulting logic by \MDN.
Serial monotone neighbourhood frames underlie the semantics of game
logic~\cite{Parikh83},
discussed in more detail in Section~\ref{sec:tboxes}. Seriality means
that each state has some neighbourhood and the empty set is never a
neighbourhood.  Notice that in the mentioned encoding of monotone
modal logic into normal modal logic, serial monotone neighbourhood
frames correspond exactly to serial Kripke frames.  It is easy to see
that the set of rules $\+R_{\MDN} = \cset{K_1,K_0,D_1}$
is a complete tableau system for $\MDN$ (notice that $K_0$ and $D_1$ are just the
instances of $K_1$ for $\Box\top$ and $\Diamond\top$, respectively).

\paragraph{Coalition logic and alternating-time logics.}
  Coalition logic~\cite{Pauly02} is essentially the next-step
  fragment of the alternating-time $\mu$-calculus AMC~\cite{AlurEA02},
  discussed in Section~\ref{sec:tboxes}. A \emph{coalition} is a
  subset of a fixed set $N=\{1,\dots,n\}$ of agents and one has
  a modal operator $[C]$ for each coalition $C$. Intuitively, we
  read formula $[C]\phi$ as `coalition $C$ has a joint strategy to enforce
  that $\phi$ shall hold in the next state'. Formally, the semantics is over
  \emph{game frames}, where for each state $x$ we have a function $f_x$
  with domain $S_1 \times \dots \times S_n$, each $S_q$
  being a finite 
  set of actions 
  available to agent $q\in N$ in state $x$. Intuitively, the choice of
  an action by each agent determines a successor state as specified by
  the \emph{outcome function} $f_x$. One then defines the semantics of
  $[C]$ by putting $x\models[C]\phi$ iff there exists a joint choice
  $(s_q)_{q\in C}$ of actions for the agents in $C$ such that for each
  joint choice $(s_q)_{q\in N-C}$ for the agents outside $C$,
  $f((s_q)_{q\in N})\models \phi$.  Note that each choice of $N$
  defines a different logic $\CL_N$ (in the sense that extending
  $\CL_{N_0}$ to $\CL_{N_1}$ for $N_0 \subsetneq N_1$ does \emph{not}
  preserve subsumption), since the semantics of $[C]$ depends on how
  many agents there are  outside $C$. For a fixed $N$, the
  set of rules $\+R_{\CL_N} = \cset{C_{ij}, C'_k : i,j \geq 0, k > 0}$
  (Fig.~\ref{fig:tab}) yields a complete tableau system for
  \CL~\cite{CirsteaEA11,SchroderPattinson09a}.

\begin{figure}

  \begin{align*}
    {\textstyle K_n\ }
     &  \frac
       {\Box a_1,\dots, \Box a_n, \Diamond b}%
       {a_1,\dots, a_n,b}
    &
    {\textstyle D_n\ }
     & \frac
       {\Box a_1, \dots,\Box a_n}
       {a_1, \dots, a_n}
    \\
   {\textstyle C_{nm}\ }
    & \frac
       {[C_1]a_1,\dots, [C_n]a_n, \tup{D}b, \tup{N}c_1,\dots, \tup{N}c_m}
       {a_1,\dots, a_n, b, c_1,\dots, c_m}
       \ {\scriptstyle \dagger\ddagger}
    &
   {\textstyle C'_{n}\ }
    & \frac
       {[C_1]a_1,\dots, [C_n]a_n}
       {a_1,\dots, a_n}
       \ {\scriptstyle \dagger}
  \end{align*}
  \caption{Tableau rules, with side conditions: {\smaller ($\dagger$)} $i\neq j \Rightarrow C_i \cap C_j = \emptyset$,
  and {\smaller ($\ddagger$)} $C_i \subseteq D$.}
  \label{fig:tab}
\end{figure}

\medskip \noindent We include only
the basic definitions of coalgebraic logic, which is more
comprehensively presented
elsewhere~\cite{Pattinson03,SchroderPattinson08,SchroderPattinson11}.
The generality of the framework stems from the parametricity of its
syntax and semantics. The language depends on a \emph{similarity type}
$\Lambda$,
which may include atomic propositions, seen as modalities of arity $0$. To simplify notation,
we pretend that all modal operators are unary.
%
The grammar for the set $L(\Lambda)$ of \emph{positive $\Lambda$-formulas} is
  \begin{equation*}
    \phi,\psi::=\top \mid\bot\mid  \phi\land\psi\mid  \phi\lor\psi\mid\M\phi
    \qquad(\M\in\Lambda).
  \end{equation*}
  The set of \emph{conjunctive} $\Lambda$-formulas is obtained by
  dropping the clauses for $\bot$ and $\lor$ from the grammar above.
  When $\Lang$ is a logic, we refer to the restriction of $\Lang$ to
  conjunctive formulas as \emph{conjunctive~$\Lang$}.

  Given a modality $\M\in\Lambda$ we use $\bar\M$ to denote the
  \emph{dual} of $\M$, with $\bar\M\phi$ interpreted as
  $\neg\M\neg\phi$ (under the usual meaning of $\neg$); we also use
  $\bar\Lambda:=\{\bar\M : \M\in\Lambda\}$. We do \emph{not} assume
  that $\Lambda$ is closed under duals, as inclusion or non-inclusion
  of dual operators in $\Lambda$ usually makes a big difference for
  the existence and size of materializations
  (Section~\ref{sec:wsi-models}).

  The semantics is parametrized, first, in terms of an endofunctor $T$
  on the category $\Set$ of sets and maps, which determines the class
  of models.  For a fixed $T$, a \emph{model} is then just a
  \emph{$T$-coalgebra} $C = (X,\xi)$, consisting of a set $X$ (of
  \emph{states}) and a \emph{transition function} $\xi : X \to TX$. A
  \emph{pointed model} is a pair $(C,r)$, where $r$ is a state of $C$,
  called the \emph{point} or \emph{root}. The intuition here is that
  $\xi(x)$ is the \emph{local view} of the model standing on a state
  $x$; e.g., in a Kripke model, $\xi(w)$ would consist of the set of
  immediate successors of world $w$, plus the set of propositions that
  hold at $w$; thus, the class of all Kripke models arises as the
  class of all $T$-coalgebras for the functor $TX = \fP(X) \times
  \fP(\mathsf{Prop})$.  As usual, we assume w.l.o.g.\ that $T$ is
  non-trivial, i.e.\ $TX=\emptyset\implies X=\emptyset$ (otherwise,
  $TX=\emptyset$ for all $X$) and preserves subsets, i.e.\
  $TX\subseteq TY$ whenever $X\subseteq Y$. (This is w.l.o.g.\ as we
  can assume that $T$ preserves injective maps, possibly after
  changing $T\emptyset$ in a way that does not affect the class of
  coalgebras~\cite{Barr93}.)

The second parameter of the semantics is the interpretation of the modal operators,
which relies on associating to each $\M\in\Lambda$ a
\emph{predicate lifting} $\bbrack{\M}$, i.e.\ a natural transformation
$\bbrack{\M} : \fQ \natto \fQ\circ T^\mathit{op}$, where
$\fQ:\Set^\mathit{op} \to \Set$ is the contravariant powerset
functor.
 That is, $\fQ X = 2^X$ for every set $X$, and for a map $f$,
$Qf$ takes preimages under $f$. In particular, naturality of $\bbrack{\M}$
means that $\bbrack{\M}_X(f^{-1}[A])=(Tf)^{-1}[\bbrack{\M}_Y(A)]$ for any
map $f:X\to Y$. 

Intuitively, a predicate lifting $\bbrack{\M}$ tells us what the local view of
a state in $X$ should be for it to satisfy a formula $\M\phi$ where $\phi$ has extension $A \subseteq X$;
explicitly, the local view $\xi(x)$ of $x$ should be an element of the set $\bbrack{\M}_X(A)$.
E.g., one interprets $\Box$ on the Kripke functor $T$ above using the predicate lifting
$$
\bbrack{\Diamond}_X(A) := \cset{ (S,V) : S \subseteq A, V \in \fP(\mathsf{Prop}) }\, .
$$
Formally, the notion of \emph{satisfaction} of
$\Lambda$-formulas $\phi$ at states $x$ of $C$ (denoted
$x\models_C\phi$) is then defined by the expected clauses for Boolean
operators, plus:
\begin{equation*}
  x \models_C \M\phi \iff \xi(x) \models \M\bbrack{\phi}_C
\end{equation*}
where $\bbrack{\phi}_C = \cset{x \in X : x \models_C \phi}$ is
the \emph{extension} of $\phi$ in $C$, and, for $t \in TX$ and $A
\subseteq X$, $$t \models \M A$$ is a more suggestive notation for $t
\in \bbrack{\M}_X(A)$. 
%
From $\Sem{\M}$ we obtain the predicate lifting interpreting
$\bar\M$ by $\Sem{\bar\M}_X(A)=TX-\Sem{\M}_X(X-A)$.

On positive formulas, the core reasoning task is \emph{subsumption checking}:
for formulas $\phi$ and $\psi$, we
say that \emph{$\psi$ subsumes $\phi$}, and write
$\phi\sqsubseteq\psi$, if $\Sem{\phi}_C\subseteq\Sem{\psi}_C$ in all
$T$-coalgebras~$C$.

Abusing notation, we identify a similarity type $\Lambda$ with
this semantic structure $\tup{T,\bbrack{\M}_{\M \in \Lambda}}$ used
to interpret it, and refer to both as $\Lambda$.
We shall use $T$ for the underlying functor throughout.


\begin{example}\label{expl:graded}
  All logics discussed above are coalgebraic; see,
  e.g.,~\cite{SchroderPattinson11,GorinSchroder13}. As an additional
  example, \emph{graded (modal) logic}, which we call $G$, has the
  similarity type $\Lambda = \cset{\Diamond_k : k \in \mathbb{N}}$,
  with $\Diamond_k\phi$ read `$\phi$ holds in more than $k$
  successors', and is interpreted over the multiset functor $\fB$,
  i.e., $\fB X = X \to \mathbb{N} \cup \cset{\infty} $. We regard
  $b\in\fB X$ as an $\NN\cup\{\infty\}$-valued measure on $X$, and
  correspondingly write $b(A)=\sum_{x \in A}b(x)$ for any subset
  $A\subseteq X$ (then, for a map $f$, $\fB f$ acts by taking image
  measures, i.e.\ $\fB f(\mu)(y)=\mu(f^{-1}[\{y\}]$.) Coalgebras for
  $\fB$ are \emph{multigraphs}, i.e.\ directed graphs whose edges are
  annotated with multiplicities from $\NN\cup\{\infty\}$. Each
  $\Diamond_k$ is interpreted by the predicate lifting
 $$  \bbrack{\Diamond_k}_X(A) := \cset{b \in \fB X : b(A) > k}.$$
 A multigraph $(X,\xi)$ is essentially a more concise representation
 of a Kripke frame, with $\xi(x)(y)=n$ standing for $n$ distinct
 successors of $x$, all of them isomorphic copies of $y$. Thus,
 $\bbrack{\Diamond_k}$ clearly captures the informal reading
 of~$\Diamond_k$.
\end{example}

\noindent
This framework is modular~\cite{SchroderPattinson11}, and in
particular supports fusion of modal logics by taking products of
functors.  For instance, the functor inducing Kripke models with $m$
relations, supporting the interpretation of $m$ relational modalities,
can be seen as arising from the product $TX=\prod_{i=1}^m\Pow(X)\times
2^{\mathsf{Prop}}$ of $m$ copies of the covariant powerset functor
$\fP$, and a copy of the constant functor $2$ given by
$2X=2=\cset{0,1}$ for each proposition symbol in $\mathsf{Prop}$
(the associated predicate liftings are derived in the obvious way).

Although coalgebraic logic supports non-monotone modalities, we assume
operators to be \emph{monotone} ($A \subseteq B \subseteq X
\Rightarrow \bbrack{\M}_X A \subseteq \bbrack{\M}_X B$): to
characterize formulas by simulations, we need monotonicity in
inductive proofs, since simulations preserve but do not reflect
satisfaction of formulas. Crucially, all monotone coalgebraic logics
admit complete sets of tableau rules consisting (besides the standard
propositional rules) of rules of the form $\Gamma_0/\Gamma_1\mid \dots
\mid\Gamma_n$ where $\Gamma_0$ contains only formulas $\M a$, with
$\M\in\Lambda\cup\bar\Lambda$, and $\Gamma_1,\dots,\Gamma_n$ contain
only variables, as in Fig.~\ref{fig:tab}~\cite{CirsteaEA11}; we fix
such a rule set $\Rules$ throughout.

In coalgebraic logic one exploits locality and reduces logical phenomena such
as derivability or satisfiability from the full logic to the simpler
setting of \emph{one-step models}, which are, roughly, the result of
forgetting the structure of a pointed model everywhere except at
the root; see, e.g., \cite{SchroderPattinson08}.  With one-step models
come \emph{one-step formulas}, i.e.\ shallow modal formulas where
propositional variables are introduced as placeholders for complex
argument formulas under modal operators.

\begin{defn}[One-step logic]\label{def:osm}\rm
  Let $\Var$ be a set of propositional variables (not fixed, and
  typically finite); a \emph{one-step model over $V$} is just a tuple
  $(X,\tau,t)$ where $X$ is a set (possibly empty), $\tau : \Var \to
  \fP X$ interprets propositional variables, and $t \in TX$. The dual
  representation of $\tau$ is $\breve{\tau}: X \to \fP\Var$, i.e.\
  $\breve{\tau}(x)= \cset{p : x \in \tau(p)}$.  A \emph{conjunctive
    one-step $\Lambda$-formula} is a finite conjunction of atoms $\M
  p$, where $\M\in\Lambda$, $p \in\Var$. The satisfaction relation is
  given by $(X,\tau,t) \models_\tau \Land_{i\in I} \M_i p_i$ iff $t
  \models \M_i\tau(p_i)$ for all~$i$.  Similarly, a \emph{positive
    one-step $\Lambda$-formula} is an element of
  $\Pos(\Lambda(\Pos(V)))$, where $\Lambda(W)=\{\M w : \M\in\Lambda,
  w\in W\}$ and $\Pos$ denotes positive propositional combinations
  (using $\top$, $\bot$, $\lor$, $\land$), with the expected
  semantics. We write $\ossub$ for the subsumption relation in the
  one-step logic: $\phi\ossub\psi$ if $(X,\tau,t)\models\psi$ whenever
  $(X,\tau,t)\models\phi$.
\end{defn}

\noindent
The transfer of results between the one-step and the full logic is
done by way \emph{collages}, i.e., pasting pointed coalgebras into a
one-step model to form a new coalgebra, and
\emph{d\'ecollages}, tearing away most of the structure of a pointed
coalgebra to obtain a one-step model (see e.g. the construction
of shallow models in~\cite{SchroderPattinson08,MyersEA09}).
Explicitly:

\begin{defn}\rm\label{def:collage}
  Given $t\in TX$, a family of pairwise disjoint pointed coalgebras
  $(C_x,x) = ((Y_x,\xi_x),x)$ for all $x\in X$, and a fresh root state
  $r$, the \emph{collage} of these \emph{collage data} is the pointed coalgebra
  $(C,r)$, with $C = (Y,\xi)$, where $Y$ is the (disjoint) union of
  $\cset{r}$ and the $Y_x$, and
          $$
    \xi(y) :=
      \begin{cases}
          t & \text{if $y = r$}
          \\
          \xi_x(y)) & \text{otherwise, for the $x$ such that $y \in Y_x$}
      \end{cases}
          $$
    \noindent
    As indicated earlier, we assume that $T$ preserves subsets,
    so, e.g., $TX\subseteq TY$.
\end{defn}

\noindent
In a nutshell, the collage is obtained from a root state $r$ with
successor structure $t\in TX$ by replacing every $x\in X$ with a pointed
coalgebra $(C_x, x)$.
The following is immediate by construction:
\begin{lemma}[Collage lemma]\label{lem:collage}
  For a collage $(C,r)$ with collage data as in
  Definition~\ref{def:collage}, and all $x \in X$, $A \subseteq Y$ and
  $\M \in \Lambda$,
    \begin{enumerate}
      \item $x \models_C \phi \iff  x \models_{C_x} \phi$, and
      \item $t \in \M_X (A \cap X) \iff \xi(r) \in \M_Y A$.
    \end{enumerate}
\end{lemma}
\begin{pf}
  The second equivalence follows directly from naturality of $\M$. For the first one, one proceeds
  by induction on $\phi$; the relevant case is the modal one:
   \begin{align*}
    x \models_\xi \M\psi
         &\iff T(\into_{Y_x})(\xi_x(x)) \in \M_Y\bbrack{\psi}_\xi
       \\&\iff \xi_x(x) \in \M_{Y_x}(\bbrack{\psi}_\xi \cap Y_x)  &&\text{(naturality)}
       \\&\iff \xi_x(x) \in \M_{Y_x}\bbrack{\psi}_{\xi_x} &&\text{(IH)}
       \\&\iff x \models_{\xi_x} \M\psi &&\hfill\qed
   \end{align*}
\end{pf}
\noindent One typically needs collages based on interpretations of
propositional variables as modal formulas. Here, we will be interested
in \emph{preserving} the interpretation of the satisfied
atoms
; more precisely:
\begin{defn}\rm
  Given collage data as in Definition~\ref{def:collage}, a valuation
  $\tau:V\to\Pow(X)$ \emph{(positively) matches} a substitution
  $\rho : \Var \to L(\Lambda)$ if for all $x \in X$, $x \models_{C_x}
  \rho(p)$ iff (if) $x\in\tau(p)$.
\end{defn}
\begin{lemma}\label{lem:fulfilling-collage}
  Let $\tau:V\to\Pow(X)$ \emph{(positively) match} $\rho : \Var \to
  L(\Lambda)$. Then
  \begin{enumerate}
  \item $x\in\tau(p)$ iff (implies) $x \models_C \rho(p)$, and
    \item $t \models_\tau \M p$ iff (implies) $r \models_C \M\rho(p)$.
  \end{enumerate}
\end{lemma}
\noindent The converse process is as follows.
\begin{defn}\rm
  Given a pointed coalgebra $(C,r)$ with $C = (X,\xi)$ and a
  substitution $\rho : \Var \to L(\Lambda)$,
  we say that $(X,\tau,t)$ is the \emph{d\'ecollage of $(C,r)$ by $\rho$} if $t = \xi(r)$ and
  $\tau(p) = \bbrack{\rho(p)}_C$.
\end{defn}

\begin{lemma}[D\'ecollage lemma]
  If $(X,\tau,t)$ is a d\'ecollage of $(C,r)$ by $\rho:\Var\to
  L(\Lambda)$ then for all one-step formulas $\phi$ over $\Var$ we
  have $(X,\tau,t) \models \phi \iff r \models_C \phi\rho$.
\end{lemma}

\section{Coalgebraic Simulations}\label{sec:coalg-el}

\noindent We recall the notion of coalgebraic modal
simulation from~\cite{GorinSchroder13}.
Given a binary relation $S\subseteq X\times Y$, we denote by $S^-$
its relational inverse. Moreover, for $A\subseteq X$,
the relational image of $S$ over $A$ is given by
$S[A]:=\{y : \exists x\in A.\,xSy\}$.

\begin{defn}[$\Lambda$-Simulation]\rm
  Let $C=(X,\xi)$ and $D=(Y,\zeta)$ be two given $T$-coalgebras. A
  \emph{$\Lambda$-simulation} $S:C\to D$ (\emph{of $C$ by $D$}) is a
  relation $S\subseteq X\times Y$ such that $xSy$ and
  $\xi(x)\models \M A$ imply $\zeta(y)\models \M S[A]$, for
  all $\M\in\Lambda$ and $A\subseteq X$.
  When $xSy$ for a $\Lambda$-simulation $S$, we say that \emph{$(D,y)$
  $\Lambda$-simulates $(C,x)$}.
\end{defn}
\noindent The properties of $\Lambda$-simulations that we need
here are the following (cf.~\cite{GorinSchroder13}):
\begin{lemma}\label{lem:comp}
  $\Lambda$-simulations are stable under relational composition;
  moreover, (graphs of) identities are $\Lambda$-simulations.
\end{lemma}

\begin{lemma}\label{lem:sim-preserves-truth}
  Let $S:C\to D$ be a $\Lambda$-simulation and $\phi$ be a positive
  $\Lambda$-formula. Then $xSy$ and $x\models_C\phi$ imply
  $y\models_D\phi$.
\end{lemma}%
\noindent The effect of dualizing modal operators is to turn around the notion
of simulation:
\begin{proposition}\label{prop:dual}
  Let $\bar\Lambda:=\{\bar\M : \M\in\Lambda\}$. A relation $S$
  between $T$-coalgebras is a $\bar\Lambda$-simulation iff $S^-$ is a
  $\Lambda$-simulation.
\end{proposition}
\begin{example}\label{expl:sim}
  (See~\cite{GorinSchroder13} for details.)
  \begin{enumerate}
  \item Over Kripke frames and for  $\Lambda=\{\Diamond\}$, a
    $\Lambda$-simulation $S:C\to D$ is just a simulation $C\to D$ in
    the usual sense.  By Proposition~\ref{prop:dual}, for
    $\Lambda=\{\Box\}$, a $\Lambda$-simulation $S:C\to D$ is then
    a simulation $D\to C$ in the usual sense.  Consequently, a
    $\{\Box,\Diamond\}$-simulation is just a standard bisimulation.
  \item A $\{p\}$-simulation for a proposition $p$ is just a
    relation that preserves $p$.
  \item Over monotone neighbourhoods with $\Lambda=\{\Box\}$,
    $S\subseteq X\times Y$  is a
    $\Lambda$-simulation between
    $\fmN$-coalgebras $(X,\xi)$ and $(Y,\zeta)$ iff
    $xSy$ and $A\in\xi(x)$ imply $S[A]\in\zeta(y)$.
  \end{enumerate}
\end{example}

\section{Weakly Simulation-Initial Models}\label{sec:wsi-models}

\noindent
In general, modal formulas need not have smallest models under the
simulation preorder. In some cases, however, such smallest models do
exist. Formally, we define this property as follows.

\begin{defn}[wsi models]\rm
  Let $\phi$ be a positive $\Lambda$-formula. A pointed
  model $(C_\phi,x_\phi)$ is called \emph{weakly simulation-initial (wsi) for
  $\phi$} if for any other 
   $(D,y)$, $y\models_D\phi$ iff
  $(D,y)$ $\Lambda$-simulates $(C_\phi,x_\phi)$.
\end{defn}

\noindent
In the relational setting, the term \emph{sim-initial} has been used
for an analogous notion~\cite{LutzWolter12}. Initiality in this sense
is rather weak, though, since the witnessing simulations are not necessarily unique.

\begin{remark}\label{rem:canonical-alt}
  Since identities are $\Lambda$-simulations, a wsi model for
  $\phi$ satisfies $\phi$. Thus, by
  Lemma~\ref{lem:sim-preserves-truth}, $(C_\phi,x_\phi)$ is
  wsi for $\phi$ iff (i) $x_\phi\models\phi$, and (ii)
  whenever $(D,y)$ is such that $y\models_D\phi$,
  then $(D,y)$ $\Lambda$-simulates $(C_\phi,x_\phi)$.
\end{remark}
\begin{defn}[\cite{LutzEA10}]\rm
  A pointed coalgebra $(C,x)$ is a \emph{materialization} of $\phi$ if
  for all positive $\Lambda$-formulas $\psi$, ${x\models_C\psi}$ iff
  $\phi\sqsubseteq\psi$. In this case, $\phi$ is
  \emph{materializable}.
\end{defn}
\noindent Of course, this definition implies that a materialization of
$\phi$ is a model of $\phi$. By Lemma~\ref{lem:sim-preserves-truth},
the following is immediate:
\begin{lemma}\label{lem:min}
  Every wsi model is a materialization.\hfill\qed
\end{lemma}

\noindent Thus, subsumption reduces to model checking in wsi models when they exist.
\begin{defn}[Convexity]\rm \cite{BaaderBrandtLutz05} A satisfiable
  $\Lambda$-formula $\phi$ is \emph{(strongly) convex} if whenever
  $\phi\sqsubseteq\Lor_{i\in I}\psi_i$ for some (possibly infinite)
  index set $I$ and positive $\Lambda$-formulas $\psi_i$ (with the
  expected semantics of $\Lor$), then already
  $\phi\sqsubseteq\psi_i$ for some $i\in I$.
\end{defn}
\begin{lemma}\label{lem:convex}
  If $\phi$ is materializable then $\phi$ is strongly convex.\qed
\end{lemma}
\begin{remark}\label{rem:convex}
  Convexity is generally felt to be necessary for tractability; see,
  e.g.,~\cite{KrisnadhiLutz07,LutzWolter12} (where it is
  considered w.r.t.\ \emph{finite} disjunctions).  It is not only an
  important structural property but also provides a good
  handle for showing that certain formulas are \emph{not} materializable.
   E.g.\ a formula that is itself a disjunction can have
  a materialization only when it is equivalent to one of its disjuncts.  It
  is thus no surprise that tractable logics such
  as \EL and TBox-free \FL  exclude disjunction; also
  here, we will henceforth \emph{restrict attention to conjunctive
    formulas}.

  But even conjunctive formulas may fail to be materializable.
  E.g., in $G$ with $\Lambda=\{\Diamond_k : k\in\NN\}$
  we have
   $\Diamond_1 a\land \Diamond_1 b \sqsubseteq \Diamond_2(a\lor b) \lor
    \Diamond_1(a\land b)$
    but the left hand side is not subsumed by any of the disjuncts of
    the right hand side, so convexity fails (cf.~\cite{BaaderBrandtLutz05}).
    Similarly, conjunctive $\{\Box_1\}$-formulas may fail to be convex, as
    witnessed by
  \begin{multline*}
    \Box_1(a\land b)\land\Box_1(b\land c)\land\Box_1(c\land d)\land\Box_1(d\land a)\sqsubseteq\\\Box_1(a\land b\land c)\lor\Box_1(b\land c\land d )\lor\Box_1(c\land d\land a)\lor\Box_1(d\land a\land b)\,.
  \end{multline*}
%
  Worse, with the wrong choice of $\Lambda$, even $\top$ may fail to
  be materializable: in $K$ with $\Lambda=\{\Box,\Diamond\}$
  we have
  $\top\sqsubseteq\Box\Diamond\top\lor\Diamond\top$ but
  $\top\not\sqsubseteq\Box\Diamond\top$ and
  $\top\not\sqsubseteq\Box\Diamond\top$.
  Similarly, in $M$ with
  $\Lambda=\{\Box,\Diamond\}$, one has that
  $\top\sqsubseteq\Box\top\lor\Diamond\top$ and yet
  $\top\not\sqsubseteq\Box\top$ and $\top\not\sqsubseteq\Diamond\top$.
\end{remark}

\noindent The existence of wsi models thus depends strongly on the
chosen $T$-structure~$\Lambda$, as well as on slight variations in the
semantics (e.g.\ w.r.t.\ seriality).  We now proceed to show that one
can limit the study of the phenomenon to the level of the much simpler
one-step logic (Section~\ref{sec:prelim}). As suggested by
Remark~\ref{rem:canonical-alt}, we define in this case:

\begin{defn}[one-step wsi models]\label{def:os-univ-sim}\rm
  A one-step model $(X,\tau,t)$ is \emph{weakly simulation-initial (wsi)}
    for a conjunctive one-step formula $\phi$ over $\Var$ if
    (i) $t \models_\tau \phi$, and
    (ii) for every 
    $(Y,\theta,s)$,
     $A \subseteq X$ and $\M \in \Lambda$,
    $t \in \M_XA$ implies $s \in \M_Y S[A]$,
  where  $xSy\iff\breve{\tau}(x)\subseteq\breve\theta(y)$.
\end{defn}

\begin{remark}\label{rem:os-univ}
  One-step wsi models are never unique.
  However, one can assume w.l.o.g.\ that if
  $(X,\tau,t)$ is wsi, then every $x\in X$ is uniquely determined by
  $\breve{\tau}(x)$ (quotient $(X,\tau,t)$ by the equivalence relation
  induced by $\breve{\tau}$), and hence that $X$ is of at most
  exponential size on the number of variables.
\end{remark}

\begin{defn}\rm
  We say that $\Lambda$ \emph{admits (one-step) wsi models}
  if every conjunctive (one-step) formula has a (one-step) wsi
  model.
\end{defn}

\noindent The main technical result of this section is then the
following.

\begin{theorem}\label{thm:univ}
  $\Lambda$ admits wsi models whenever it
  admits one-step wsi models.
\end{theorem}




\begin{proof*}{Proof (Sketch)}
  Induction on $\phi$. We have $\phi = \Land_{i \in I}\M_i\chi_i$
  for a finite (possibly empty) set $I$.  Take $\Var_\phi :=
  \cset{a_{\chi_i} : i \in I}$ and decompose $\phi$ as
  $\phi=\phi^*\rho$ with $\phi^* := \Land_{i\in I}\M_i a_{\chi_i}$ a
  one-step formula and $\rho(a_{\chi_i}):= \chi_i$ a substitution.
  Let $(X,\tau,t)$ be a wsi for $\phi^*$. By
  IH, there is, for each $x \in X$, a wsi model $(C_x,x)$
  with $C_x=(Y_x,\xi_x)$ for $\Land_{p \in\breve\tau(x)}\rho(p)$ with
  root $x$; the $Y_x$ can be assumed pairwise disjoint. Pick a fresh $x_\phi$,
  and obtain $(C_\phi,x_\phi)$ by taking $\xi(x_\phi)=t$ and attaching
  $C_x$ at each $x\in X$ (cf.~¸\cite{SchroderPattinson08}). One easily
  shows that $(C_\phi,x_\phi)$ is wsi for $\phi$.
\end{proof*}
\noindent We now analyze under which conditions one-step
wsi models exist. To begin, note that at the one-step
level, wsi models coincide with materializations
(recall that $\ossub$ is the one-step subsumption relation of Def.~\ref{def:osm}):
\begin{defn}\rm
  A one-step model $(X,\tau,t)$ is a \emph{one-step materialization}
  of a conjunctive one-step $\Lambda$-formula $\phi$ over $V$ if for
  every literal $\hearts\rho$ with $\hearts\in\Lambda$ and
  $\rho\in\Pos(V)$, $t\models_\tau \hearts\rho$ iff
  $\phi\ossub\hearts\rho$. In this case, $\phi$ is
  \emph{materializable}.
\end{defn}
\noindent
Again, this implies that a one-step materialization of $\phi$
is a model of $\phi$.
\begin{lemma}\label{lem:os-mat}
  A one-step model is wsi for a conjunctive one-step $\Lambda$-formula
  $\phi$ iff it is a materialization of $\phi$.
\end{lemma}
\noindent Moreover, existence of materializations is equivalent to convexity:
\begin{defn}\rm\label{def:os-convex}
  A satisfiable one-step $\Lambda$-formula $\phi$ over $V$ is
  \emph{strongly convex} if whenever $\phi\ossub\Lor_{i\in I}\psi_i$
  for positive one-step $\Lambda$-formulas $\psi_i$ over $V$ and a
  (possibly infinite) index set $I$, then already $\phi\ossub\psi_i$
  for some $i\in I$.
\end{defn}
\begin{remark}
  In case $\Lambda$ is finite, strong convexity of one-step formulas
  is the same as convexity (the notion obtained by restricting $I$ to
  be finite in Definition~\ref{def:os-convex}), as then there are,
  up to equivalence, only finitely many positive one-step
  $\Lambda$-formulas over~$V$.
\end{remark}
\begin{lemma}\label{lem:os-convex}
  A one-step $\Lambda$-formula is materializable iff it is strongly
  convex.
\end{lemma}
\begin{remark}
  Summing up, at the one-step level the notions of \emph{being
    materializable}, \emph{having a wsi model} and \emph{being
  strongly convex} coincide. For the full logic, we have already
  noted that wsi models are materializations and materializable formulas
  are strongly convex. We leave the equivalence of these
  notions for individual formulas, i.e.\ to show that every strongly
  convex formula has a wsi model, to future research (for
  some relational logics, this equivalence is
  known~\cite{AcetoEA11,BoudolLarsen92}). Under mild additional
  assumptions, it does follow at the current stage that the
  equivalence holds between the respective properties of the logic as
  a whole: assume for simplicity that $\Lambda$ contains infinitely
  many proposition symbols (actually, it suffices that the logic is
  \emph{non-trivial}, i.e.\ contains infinitely many propositionally
  independent formulas). If all conjunctive $\Lambda$-formulas are
  strongly convex, then this holds (emulating propositional variables
  by proposition symbols from $\Lambda$) also for conjunctive one-step
  $\Lambda$-formulas. By the above, it follows that $\Lambda$ admits
  one-step wsi models, and hence admits wsi models.
\end{remark}

\noindent Next, we show how to read off convexity from the structure
of the tableau rules for $\Lambda$. At the same time, we obtain
a description of the structure of one-step materializations.
\begin{defn}
  \rm We call a tableau rule \emph{definite} if it has exactly one
  conclusion, i.e.\ is of the form $\Gamma/\Delta$ with
  $\Gamma\subseteq(\Lambda\cup\bar\Lambda)(V)$ and $\Delta\subseteq
  V$. A set $\Rules$ of definite one-step rules \emph{preserves
    $\Lambda$-convexity} if whenever a rule $R$ over $V$ in $\Rules$
  can be written in the form
  $\Gamma_1,\Gamma_2/\Delta_1,\Delta_2\in\Rules$ with
  $\Gamma_1\subseteq\Lambda(V_1)$,
  $\Gamma_2\subseteq\bar\Lambda(V_2)$, $\Delta_1\subseteq V_1$,
  $\Delta_2\subseteq V_2$, with $V_1$, $V_2$ a disjoint decomposition
  of $V$ (we call this a \emph{$\Lambda$-splitting} of $R$), then for
  each $\bar\hearts a\in\Gamma_2$, the rule $\Gamma_1,\bar\hearts
  a/\Delta_1,a$ is also in $\Rules$.
\end{defn}
The next theorem will show that preservation of $\Lambda$-convexity is
sufficient for convexity of conjunctive $\Lambda$-formulas. It is
fairly clear that, in cases where  all rules are definite, necessity also
holds for a sufficiently carefully formulated weakening of
preservation of $\Lambda$-convexity (e.g.\ in the above notation, it
clearly suffices to have $\Gamma_1,\bar\hearts a/\Delta_1,a$ derivable
from $\Rules$ in the obvious sense); we refrain from exploring
details.
\begin{remark}\label{rem:convex-dual}
  In case $\Lambda$ is closed under duals, 
  the rule set $\Rules$ preserves $\Lambda$-convexity iff whenever
  $\Gamma/\Delta$ is a rule over $V$ in $\Rules$ and $\emptyset\neq
  V_0\subseteq V$, then $(\Gamma\cap\Lambda(V_0))/(\Delta\cap V_0)$ is
  in $\Rules$ -- that is, iff $\Rules$ is stable under deleting
   variables.
\end{remark}
\begin{theorem}\label{thm:convex-rules}
  Let $\Lambda$ be finite (for brevity; in fact it suffices to assume
  a more sophisticated form of
  completeness~\cite{SchroderPattinson09b}). If $\Rules$ preserves
  $\Lambda$-convexity, then $\Lambda$ admits wsi models.
  Moreover, a one-step materialization for a conjunctive
  one-step $\Lambda$-formula $\phi=\Land_{i\in I}\M_ia_i$ (read also
  as the set $\{\M_i a_i : i\in I\}$) is then obtained as
  follows. First put $W=\{a_i: i\in I\}$, and define $(X,\tau)$ to
  consist of
  \begin{itemize}
  \item a state $x$ with $\breve\tau(x)=\Delta\sigma$, for each rule
    $\Gamma/\Delta$ over $V$ in $\Rules$ and each renaming
    $\sigma:V\to W$ with $\Gamma\sigma\subseteq\phi$;
  \item a state $x$ with $\breve\tau(x)=\Delta_1\sigma$, for each rule
    $\Gamma,\bar\M b/\Delta_1,\Delta_2$ over $V\uplus\{b\}$ in $\Rules$ with
    $\Delta_2\subseteq\{b\}$ and each renaming $\sigma:V\to W$ with
    $\Gamma\sigma\subseteq\phi$.
  \end{itemize}
  (In both cases, we can restrict to rules and renamings for which
  $\Gamma\sigma$ becomes maximal.)  Then there exists $t\in TX$ such
  that $(X,\tau,t)$ is a materialization of~$\phi$.
\end{theorem}

\begin{remark}
  The rule sets in all examples are built in such a way that $\sigma$
  can be restricted to be injective in the construction of
  Theorem~\ref{thm:convex-rules}~\cite{SchroderPattinson09a}; however,
  it is easy to see that in such cases, this restriction does not
  actually affect the result of the construction.
\end{remark}

\begin{example}\label{expl:univ-prop}
  Over the proposition functor $2$, $\Lambda=\cset{p}$ and
  $\Lambda=\cset{\bar p}$ (but not, of course,
  $\Lambda=\cset{p,\overline{p}}$) are easily seen to admit wsi models;
  e.g.\ $(\emptyset,\emptyset,1)$ is wsi for $p$. This is our only positive
  example not matching Theorem~\ref{thm:convex-rules}: the one-step
  rule $p,\bar p/\bot$ fails to be definite, having no conclusion.
\end{example}

\begin{figure}
 \centering
 \subfigure[$K$: \smaller{$\Diamond p \land \Diamond q \land \Diamond r$}]{
  \label{subfig:K-diam}
  \begin{tikzpicture}
     \node
       [root]
       (root) {};

     \node
       [n, below left of=root,label=below:{\smaller $p$}]
       (p) {};

     \node
       [n, below of=root,label=below:{\smaller $q$}]
       (q) {};

     \node
       [n, below right of=root,label=below:{\smaller $r$}]
       (r) {};

       \draw [arr] (root) to (p);
       \draw [arr] (root) to (q);
       \draw [arr] (root) to (r);
  \end{tikzpicture}
 }
~~~
 \subfigure[\KD: \smaller{$\Diamond p \land \Diamond q \land \Box r\land \Box s$}]{
   \label{subfig:KD}
   \begin{tikzpicture}
     \node
       [root]
       (root) {};

     \node
       [n, below left of=root,label=below:{\smaller $p,r,s$}]
       (p) {};

     \node
       [n, below of=root,label=below:{\smaller $r,s$}]
       (top) {};

     \node
       [n, below right of=root,label=below:{\smaller $q,r,s$}]
       (q) {};

       \draw [arr] (root) to (p);
       \draw [arr] (root) to (q);
       \draw [arr] (root) to (top);
   \end{tikzpicture}
  }
~~~
 \subfigure[\MDN: \smaller{$\Box p\land\Box q\land\Diamond r\land \Diamond s$}]{
  \label{subfig:MDN}
  \begin{tikzpicture}
    [nb/.style={nbhd
               ,minimum width=25pt
               ,minimum height=35pt}
    ]
     \node
       [root]
       (root) {};

      \node
        [nb,below left of=root
        ]
        (np) {};
     \node
       [n,below left of=root,label=left:{\smaller $p$}
       ,xshift=7pt]
       (p) {};
     \node
       [n,below left of=root,label=left:{\smaller $r,p$}
       ,xshift=7pt,yshift=10pt]
       (pr) {};
     \node
       [n,below left of=root,label=left:{\smaller $s,p$}
       ,xshift=7pt,yshift=-10pt]
       (ps) {};

     \node
        [nb,below of=root
        ,yshift=3pt
        ]
        (ntop) {};
     \node
       [n,below of=root
       ,yshift=2pt]
       (top) {};
     \node
       [n,below of=root,yshift=12pt,label=right:{\smaller $r$}]
       (r) {};
     \node
       [n,below of=root,yshift=-8pt,label=right:{\smaller $s$}]
       (s) {};

      \node
        [nb,below right of=root
        ]
        (nq) {};
     \node
       [n,below right of=root,label=right:{\smaller $q$}
       ,xshift=-7pt]
       (q) {};
     \node
       [n,below right of=root,label=right:{\smaller $q,r$}
       ,xshift=-7,yshift=10pt]
       (qr) {};
     \node
       [n,below right of=root,label=right:{\smaller $q,s$}
       ,xshift=-7,yshift=-10pt]
       (qs) {};

       \draw [arr] (root) to (np);
       \draw [arr] (root) to (nq);
       \draw [arr] (root) to (ntop);

  \end{tikzpicture}
 }


\caption{
  One-step wsi models for the indicated formulas.
  The white node is the implicit root, the black ones its domain. For \MDN, minimal
  neighbourhoods are depicted (dotted boxes), not their supersets.
  \label{fig:univ-sims}
}
\end{figure}
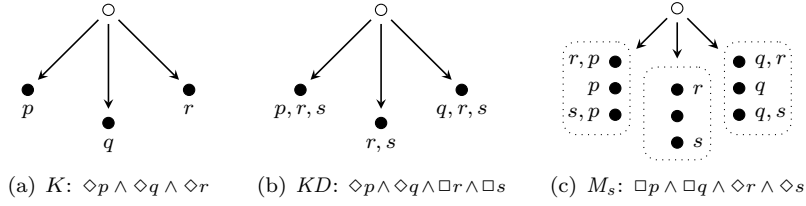

\begin{example}\label{expl:univ-K}
  Over Kripke frames, we have the following.
  \begin{enumerate}
  \item $\Lambda=\{\Diamond\}$ admits wsi models: a
    $\{\Diamond\}$-splitting $\Gamma_1,\Gamma_2/\Delta_1,\Delta_2$ of
    $(K_n)$ in Fig.~\ref{fig:tab} is of the form $\Gamma_1=\Diamond b$,
    $\Gamma_2=\Box a_1,\dots,\Box a_n$, and for each $j$ we have a
    rule $\Diamond b,\Box a_j/b,a_j$ in $\Rules_K$, as required. The
    one-step wsi model for $\Land_{i\in I}\Diamond a_i$
    according to Theorem~\ref{thm:convex-rules} is $(I,\tau,I)$ with
    $\tau(a_i)=\{i\}$. An example is depicted in
    Fig.~\ref{subfig:K-diam}. This extends to the multimodal case (see
    Remark~\ref{rem:univ-fusion}), essentially, to \EL.

  \item $\Lambda=\{\Box\}$ admits wsi models: any
    $\{\Box\}$-splitting $\Gamma_1,\Gamma_2/\Delta_1,\Delta_2$ of
    $K_n$ already has $\Gamma_2$ of the form $\Diamond b$. Restricting
    to maximal rule matches, the one-step wsi model for
    $\Land_{i\in I}\Box a_i$ according to (the second clause of)
    Theorem~\ref{thm:convex-rules} is $(\{*\},\tau,\{*\})$ with
    $\tau(a_i)=\{*\}$ for all $i$.  This extends straightforwardly to
    the multi-modal case (Remark~\ref{rem:univ-fusion}), of which
    \FL is a syntactic variant.
  \item In $K$, $\Lambda=\{\Box,\Diamond\}$ fails to be convex
    (Remark~\ref{rem:convex}). Note that $\Rules_K$ fails to preserve
    convexity: deleting $b$ from $K_n$ yields $D_n\notin\Rules_K$
    (Fig.~\ref{fig:tab}, Remark~\ref{rem:convex-dual}). In \KD,
    however, $\{\Box,\Diamond\}$ does admit wsi models, as
    the rules $K_n$ and $D_n$ together are stable under deleting
    occurrences of variables. Restricting to maximal matches, the
    one-step wsi model for $\Land_{i\in I}\Box
    a_i\land\Land_{j\in J}\Diamond b_j$ 
     is $(J\cup\{*\},\tau,J\cup\{*\})$
    (with $*\notin J$) given by $\tau(a_i)=J\cup\{*\}$ and
    $\tau(b_j)=\{j\}$ (see Fig.~\ref{subfig:KD}).
  \end{enumerate}
\end{example}


\begin{example}\label{expl:univ-monot} Over monotone neighbourhoods,
  the situation is analogous as over Kripke frames, due
  to the similarity of the rule sets: both $\Lambda=\{\Box\}$ and
  $\Lambda=\{\Diamond\}$ admit wsi models in $M$, but not
  $\Lambda=\{\Box,\Diamond\}$  ($M$ validates
  $\Box\top\lor\Diamond\top$ but none of the disjuncts, so
  no $\cset{\Box,\Diamond}$-formula is convex).  
  In \MDN, $\{\Box,\Diamond\}$ does admit wsi models, though, for
  essentially the same reasons as in \KD.  The one-step wsi model from
  Theorem~\ref{thm:convex-rules} for 
  $\Land_{i\in I}\Box a_i\land\Land_{j\in J}\Diamond b_j$ ($I$,
  $J$ disjoint) is $(X,\tau,\FN)$:
    \begin{align*}
    X & := \{K\subseteq I\cup J : |K\cap I| \le 1, |K\cap J|\le 1\}
    &
    \tau(a_i) & := \{K\in X : i\in K\}
    \\
    \FN & := {\uparrow}(\{\tau(a_i) : i\in I\}
       \cup \{\{K\in X : K\subseteq J\}\}\})
    &
    \tau(b_j) & := \{K\in X : j\in K\}
  \end{align*}
  where $\uparrow$ is closure under taking
  supersets. Fig.~\ref{subfig:MDN} depicts the construction. 
\end{example}

\begin{example}\label{expl:coal-univ}
  In coalition logic, 
  $\Lambda=\{[C],\tup{C} : C\subseteq N\}$ admits wsi models: its rules
  are stable under deleting occurrences of variables  by Remark~\ref{rem:convex-dual}.
\end{example}

  \begin{remark}\label{rem:univ-fusion}
    When one models fusion of modal logics by taking products of
    functors as noted in Sec.~\ref{sec:prelim} (see~\cite{SchroderPattinson11})
    this is
    reflected in the construction of one-step wsi models by
    just taking disjoint unions of the domains and pairing the
    transition structures (prolonged into the disjoint union). For instance,
    in the multimodal logic $\Lambda=\{\Box_1,\dots,\Box_n\}$ over Kripke frames,
    one-step wsi models for one-step formulas
    $\Land_{i=1}^n\Land_{j=1}^{m_i}\Box_i a_{ij}$  are
    formed by taking the disjoint union of the one-step wsi models for
    the formulas $\Land_{j=1}^{m_i}\Box_i a_{ij}$ as described in
    Example~\ref{expl:univ-K}, and thus have $n$ states, with the
    $i$-th state satisfying the propositional variables
    $a_{i1},\dots,a_{im_i}$. By Example~\ref{expl:univ-prop},
    adding atomic propositions does not
    enlarge the carriers of wsi models at all.
  \end{remark}

  \noindent For tractability, studied in the next section, we need
  wsi models to be small. However, existence of wsi models
  is of independent interest, even in those cases in which they may
  be exponentially large.
  For instance, from Example~\ref{expl:univ-K}, we can already
  conclude that \emph{conjunctive \KD is convex.}

  \section{Tractability}\label{sec:tractability}

  \noindent Assume from now on that $\Lambda$ admits one-step
  wsi models. Lemma~\ref{lem:min} then allows us to reduce
  subsumption to satisfaction in such models. (This is also the
  principle underlying state-of-the-art consequence-based reasoning
  procedures, which for \EL go back
  to~\cite{BaaderBrandtLutz05}.) In the previous sections, we have
  refrained from giving explicit descriptions of $t$ when $(X,\tau,t)$
  is wsi, and in fact it is not necessary to actually know $t$.
  Instead, we opt for a different
  representation of wsi models: in the recursive construction
  of a wsi model $(C_\phi,x_\phi)$ for a conjunctive
  $\Lambda$-formula $\phi$ (see proof sketch of
  Theorem~\ref{thm:univ}), we have calculated a one-step formula
  $\phi^*$ and used a one-step wsi model $(X,\tau,t)$ for
  it. For algorithmic purposes, we now drop $t$ but store $\phi^*$,
  $X$, and $\tau$; we call the arising object an \emph{abstract wsi model
  for $\phi$}. We face then the following problem:
  \begin{defn}\rm
    The \emph{conjunctive one-step consequence problem} of $\Lambda$
    is to decide, given a conjunctive one-step $\Lambda$-formula
    $\psi$ over $V$, $\hearts\in\Lambda$, and $\rho\in\Pos(V)$,
    if $\psi\ossub\hearts\rho$.
  \end{defn}
  If the conjunctive one-step consequence problem for $\Lambda$ is in
  $P$, then we can check in time polynomial in the size of an
  abstract wsi model $(C_\phi,x_\phi)$ for $\phi$ whether
  $x_\phi\models_{C_\phi}\psi$ for a positive $\Lambda$-formula
  $\psi$, e.g.\ by calculating extensions of subformulas of $\psi$
  bottom up. Now in the positive examples of the previous section,
  deciding whether, in the notation of the above definition,
  $\psi\ossub\hearts\rho$ can be done using the respective rule sets
  to check whether $\psi\land\bar\hearts\neg\rho$ is satisfiable,
  which in turn will lead to checking satisfiability of a
  propositional formula of the form $\chi\land\neg\rho$ where $\chi$
  is a conjunction over $V$, a trivial task given that $\rho$ is
  positive. Thus, the conjunctive one-step consequence problem of
  $\Lambda$ is in $P$ provided that we can polynomially bound the
  number of rule matches to a given conjunction over $\Lambda(W)$,
  which is easily seen for all relevant examples.


  Polynomial-time computability (entailing polynomially bounded size)
  of abstract wsi models will then imply tractability of
  subsumption. In some cases, tractability will hold only if we bound
  certain parameters. To avoid overformalization, we will call any set
  of conjunctive $\Lambda$-formulas a \emph{conjunctive
    $\Lambda$-fragment} and apply notions defined so far w.r.t.\ the
  set of all conjunctive formulas, such as \emph{admitting wsi models},
    also to fragments. Note that sometimes restricting to
  a fragment will also restrict the relevant set of one-step formulas.

  \begin{defn}\rm
    A conjunctive $\Lambda$-fragment $\Lang$ \emph{admits polynomial
      wsi models} if every $\Lang$-formula has a
    polynomial-time computable abstract wsi model.
  \end{defn}
  \begin{lemma}
    If $\Lang$ admits polynomial wsi models and the
    conjunctive one-step consequence problem of $\Lambda$ is in $P$,
    then subsumption $\phi\sqsubseteq\psi$ between $\Lang$-formulas
    $\phi$ and positive $\Lambda$-formulas $\psi$ is in $P$.
  \end{lemma}
  \noindent We identify tractability criteria at the one-step level:
  \begin{defn}\rm
    We say that one-step wsi models $(X,\tau,t)$ of one-step
    formulas $\phi=\Land\hearts_i a_i$ are \emph{linear} if
    $|\tau(a_i)|\le 1$ for all $i$, \emph{$k$-bounded} if
    $|\breve{\tau}(x)|\le k$ for all $x\in X$, and \emph{polynomial}
    if $|X|$ is polynomially bounded in the size of $\phi$. 
  \end{defn}

  \noindent
  In words, linearity means that every propositional variable is satisfied in
  at most one state, while $k$-boundedness means that each state satisfies
  at most~$k$ propositional variables.

  \begin{proposition}
    \textsl{a}) If a conjunctive $\Lambda$-fragment $\Lang$ admits linear
      or $k$-bounded  one-step wsi models,
      then $\Lang$ admits polynomial wsi models.
    \textsl{b}) If $\Lambda$ admits polynomial one-step wsi models,
      then  conjunctive $\Lambda$-fragments defined by bounding the
      modal depth admit polynomial wsi models.
  \end{proposition}
  \noindent (The complexity of bounded-depth fragments of modal logics
  over a complete Boolean basis has been studied, e.g.,
  in~\cite{Halpern95}.)
  \begin{proof*}{Proof (Sketch)} Linearity  implies that a
    wsi model for $\phi$ has at most as many states as $\phi$
    has subformulas. On the other hand, $k$-boundedness ensures that
    wsi models, constructed as trees in the proof of
    Theorem~\ref{thm:univ}, can be collapsed into polynomial-sized
    dags by identifying states realizing the same target formula; by
    $k$-boundedness, at most $|\phi|^k$ target formulas will arise in
    the construction.
  \end{proof*}

\begin{example}
  \begin{enumerate}
  \item One-step wsi models for $\{\Diamond\}$ and for
    $\{\Box\}$ over Kripke frames (Example~\ref{expl:univ-K}) are
    linear; those for $\{\Diamond\}$ are in addition $1$-bounded. By
    Remark~\ref{rem:univ-fusion}, this extends straightforwardly to
    the case with multiple modalities and atomic propositions. We thus
    recover the known results that subsumption checking in conjunctive
    multimodal $K$ with only diamonds (\EL) or only boxes
    ($\FL$) is in $P$. As an aside, the conjunctive
    fragment of the co-contravariant modal logic of~\cite{AcetoEA11},
    which is essentially positive Hennessy-Milner logic with only
    diamonds for some actions and only boxes for the others, can be
    seen as a fusion of a logic of boxes with a logic of diamonds, and
    thus also has linear one-step wsi models, i.e.\ has a
    polynomial-time subsumption problem.
  \item One-step wsi models for $\{\Box,\Diamond\}$ over
    serial Kripke frames, i.e.\ for conjunctive \KD, are polynomial,
    so that \emph{subsumption in bounded-depth fragments of
      conjunctive \KD is in $P$} (with unboundedly many atomic
    propositions). This may be seen as a companion result to the
    (easily proved) coNP upper bound for bounded-depth fragments of
    full $K$~\cite{Halpern95}.

    Alternatively, if one restricts conjunctive \KD formulas to use
    at most $k$ boxes at each modal depth, then one-step wsi models for
    them become $k+1$-bounded, so that this restriction also ensures
    tractable reasoning. Again, this extends easily to the multimodal
    case with unboundedly many atomic propositions. Since one has a
    straightforward embedding of \EL into multimodal \KD
    (using a fresh propositional atom $e$ marking `existing' states to
    simulate arbitrary Kripke frames with serial ones), this result
    can be seen as generalizing the tractability of \EL
    (which is just the case $k=0$).

    It is worth observing that the more specific problem of
    \emph{satisfiability} but over unrestricted conjunctive \KD
    extended with atomic negation (called \emph{poor man's logic}) is known
    to be in $P$~\cite{Hemaspaandra01}.

  \item In \MDN (Example~\ref{expl:univ-monot}), wsi models
     for $\Lambda=\{\Box,\Diamond\}$ are $2$-bounded, so that
    \emph{conjunctive \MDN is tractable}. Similarly, wsi models
    for the structure $\Lambda=\cset{[C],\tup{C} : C\subseteq
    N} - \cset{[\emptyset],\tup{N}}$ in coalition logic /
    alternating-time logic are $n$-bounded, where $n$ is the (fixed!)
    total number of agents (since $n$ is also the maximal number of
    disjoint non-empty coalitions). Thus, for each finite set $N$ of
    agents, \emph{conjunctive coalition logic over $N$ without
      $[\emptyset]$ and $\tup{N}$ is tractable}. 
  \end{enumerate}
\end{example}

\section{Greatest Fixpoints}\label{sec:tboxes}
\noindent
We now proceed to extend the base logic with a fixpoint operator. This will
allow us to cover global definitions (e.g..~classical
terminological boxes, in DL parlance) and fragments of
game logic and the alternating-time $\mu$-calculus.  We can only
expect to get wsi models for formulas with \emph{greatest}
fixpoints, which are similar in flavour to infinite conjunctions,
while least fixed points are disjunctive (e.g., $\nu x.(p \land
\Diamond x)$ can be seen as the infinitary formula $p \land \Diamond(p
\land \Diamond(p \land \dots )))$ which characterizes an infinite path
of nodes satisfying $p$).

Following~\cite{LutzEA10}, we will actually allow for mutually
recursive auxiliary definitions, as in the vectorial
$\mu$-calculus~\cite{ArnoldNiwinski01}. The resulting logic can be
shown to be no more expressive than the one with only single-variable
$\nu$, but to admit exponentially more succinct
definitions~\cite{LutzEA10}.  Syntactically, the grammar of
\emph{positive $\Lambda$-$\nu$-formulas} extends that of positive
$\Lambda$-formulas with fixpoint variables from a set $\Delta$ and,
for $\alpha\in \cset{\nu,\mu}$, formulas $\alpha(y;y_1,\dots
y_n).(\phi,\phi_1\dots\phi_n)$, where $y,y_1,\dots y_n \in \Delta$
must be distinct and $\phi,\phi_1,\dots\phi_n$ are positive
$\Lambda$-$\nu$-formulas.  A formula $\nu(y;y_1,\dots
y_n).(\phi;\phi_1,\dots, \phi_n)$ defines $y,y_1,\dots,y_n$ as a
simultaneous greatest fixpoint, and then returns $y$; similarly for
$\mu$ with least fixpoints.
  A \emph{sentence} is a formula where
every fixpoint variable is bound by a $\nu$ or $\mu$.
\emph{Conjunctive} fixpoint $\Lambda$-formulas extend conjunctive
$\Lambda$-formulas with $\nu$ only.

We define the semantics of this language over a $T$-coalgebra
$C=(X,\zeta)$ and a valuation $\+V : \Delta \to \fP(X)$; by
$\bbrack{\phi}_{C,\+V}$ we denote the extension of $\phi$ in $C$
assuming that the fixpoints variables are interpreted using $\+V$. The
propositional and modal cases are defined like before (with
$\bbrack{x}_{C,\+V} = \+V(x)$); moreover, 
$\bbrack{\nu(y_0;y_1,\dots y_n).(\phi_0;\phi_1,\dots \phi_n)}_{C,\+V}$
is the first projection of the greatest fixed point of the map taking
$(A_0,\dots,A_n)$ to $(\bbrack{\phi_i}_{C,\+V[y_0\mapsto A_0\dots y_n
  \mapsto A_n]})_{i=1,\dots,n}$. The semantics of $\mu$ is dual. For a
sentence $\phi$, the initial $\+V$ is irrelevant, so we may write just
$\bbrack{\phi}_C$.  Preservation of positive formulas by simulations
extends to fixpoint formulas:
\begin{lemma}
  Let $S$ be a $\Lambda$-simulation of a coalgebra $C=(X,\xi)$ by a
  coalgebra $D$, and let $\+V:\Delta \to \fP(X)$ be a
  valuation. Then for every positive $\Lambda$-$\nu$-formula $\phi$,
    $S[\bbrack{\phi}_{C,\+V}]\subseteq \bbrack{\phi}_{D,S[\+V]}$,
  where $S[\+V]$ denotes the valuation taking $x$ to $S[\+V(x)]$.
\end{lemma}
\noindent Extending the definition of \emph{wsi models}
literally to positive $\Lambda$-$\nu$-formulas, we thus obtain a
generalization of Lemma~\ref{lem:min}, i.e.\ a wsi model for a
fixpoint formula $\phi$ is a \emph{materialization}, so that subsumption
of $\phi$ by positive $\Lambda$-$\nu$-formulas reduces to satisfaction
in the wsi model.

\begin{example}\label{expl:tboxes}
  DLs are logics for knowledge representation, where
  terminologies are defined via axioms in \emph{TBoxes} which
  effectively constrain the classes of models over which one reasons.
  In particular, one is sometimes interested in so-called
  \emph{classical TBoxes with greatest fixpoint
    semantics}~\cite[Chapter~2]{BaaderEA07}.  Here, axioms of a TBox
  $\+T$ are definitions of the form $a \equiv \phi$ with $a$ a
  proposition symbol that is allowed to occur as a left-hand side of
  only one definition.  Such an $a$ is said to be a \emph{derived}
  concept of $\+T$. Each model $C$ interpreting the non-derived
  propositions is extended to a unique model $C^\+T$ which arises as
  the greatest fixpoint of the function mapping an extension $C'$ of
  $C$ interpreting also the derived propositions to the extension
  $C''$ where for each $a \equiv \phi \in \+T$, $\bbrack{a}_{C''} =
  \bbrack{\phi}_{C'}$. One writes $\+T \models \psi\sqsubseteq\chi$ if
  for each model $C$, $\psi\sqsubseteq\chi$ holds in $C^\+T$. It is
  then clear that \emph{subsumption over $\+T$}, i.e.\ to decide
  whether $\+T \models \psi\sqsubseteq\chi$, reduces to subsumption of
  fixpoint formulas: assume $\+T = \cset{a_1 \equiv \phi_1,\dots a_n
    \equiv \phi_n}$; we have $\+T \models \psi\sqsubseteq\chi$ iff
  $\nu(z;a_1,\dots,
  a_n).(\psi;\phi_1,\dots,\phi_n)\sqsubseteq\nu(z;a_1,\dots,
  a_n).(\chi;\phi_1,\dots,\phi_n)$ where $z$ is a fresh
  variable. Additional details are given by Lutz et
  al.~\cite{LutzEA10}.
\end{example}

\begin{exmp}[Game logic]\rm \label{ex:gl}
%
  Model-checking a PDL formula $\tup{\alpha}\top$ can be seen as
  finding a winning strategy in a one-player game, where $\alpha$
  describes the rules of the game and the model encodes the possible
  moves of the player on a fixed game board.
  In Game Logic (GL)~\cite{Parikh83}, this notion is extended
  to two-player games (of perfect information).  Composite games
  $\alpha$ are built from atomic games
  using the program constructors of PDL plus a \emph{dualization}
  operator ($\cdot^d$), which corresponds to players swapping roles,
  so that $\tup{\alpha^d}\phi\equiv[\alpha]\phi$ (and hence $[\alpha]$
  can be omitted from the
  language). 
  The two-player view disables normality (i.e.\ one no longer has
  $\tup{\alpha}(\phi\lor\psi)\to\tup{\alpha}\psi\lor\tup{\alpha}\psi$);
  hence, models of GL are products of monotone neighbourhood frames
  $S_a$,
    one per atomic game $a$.  Intuitively, a set $A
  \in S_a(x)$ corresponds to (an upper bound on) positions that could
  be reached from $x$ when following a fixed strategy for $a$;
  allowing for different responses of player II, we see that $A$ need
  not be a singleton
  . As a notational infelicity, the predicate lifting interpreting
  $\tup{a}$ in GL (for $a$ atomic) is that of $\Box$ in standard
  notation for monotone modal logic.  Serial models
  are those where atomic
  games never get stuck, no matter which player begins.
  We note that
  GL has a well-known sublogic, concurrent propositional dynamic
  logic CPDL~\cite{Peleg87}, which omits dualization
  $\cdot^d$ but retains $\cap$, the dual of $\cup$.

  GL can be embedded into the fixpoint extension $\MDN^\nu_m$ of
  multi-modal $\MDN$ (with duals of atomic propositions), much like
  PDL can be embedded into the relational $\mu$-calculus.  Two
  fixpoint variables suffice for this~\cite{BerwangerEA07}.  It is not
  hard to see that using fixpoint variables as a form of
  let-expressions, one can avoid the exponential blowup present in the
  original encoding.  The \emph{conjunctive} fragment of GL is swiftly
  defined as the preimage of the conjunctive fragment of $\MDN^\nu_m$
  under this embedding.

\end{exmp}

\begin{exmp}[Alternating time]\rm The \emph{alternating-time
    $\mu$-calculus (AMC)} is essentially the extension of coalition logic
  with fixpoint operators (its actual notation is slightly different)~\cite{AlurEA02}.
  The \emph{conjunctive fragment} of the AMC can by defined in the
  obvious way excluding $\lor$, $\neg$, and $\mu$.  In this fragment,
  we can still express `always' formulas from alternating-time
  temporal logic (ATL) such as $\ATLdiamond{C}\Box\phi$, which is read
  `coalition $C$ can maintain $\phi$ forever', and is
  equivalent to the fixpoint formula $\nu x.\,(\phi\land[C]x)$.
\end{exmp}
\noindent We proceed to show that if $\Lambda$ admits one-step
wsi models, we also obtain wsi models for
conjunctive $\Lambda$-$\nu$-formulas. We exploit the fact that any
such sentence can be put, in polynomial time, in a \emph{shallow}
normal form, i.e.\ without nested occurrences of $\nu$ (using Beki{\v
  c}'s law~\cite{ArnoldNiwinski01}) and without nesting of modal
operators (using abbreviations for subformulas in analogy to standard
TBox normalizations~\cite{BaaderIJCAI03a}).


  Thus, let $\phi =
  \nu(x_0;x_1,\dots,x_n)(\phi_0;\phi_1,\ldots,\phi_n)$ be a shallow
  sentence.  
  We shall assume, for each
  conjunctive one-step $\Lambda$-formula $\psi$ over $V = \Delta$, a
  fixed one-step wsi model $(X_\psi,\tau_\psi,t_\psi)$ which
  we then call \emph{the} one-step wsi model for $\psi$.  We
  assume w.l.o.g.\ that $X_\psi\subseteq\Pow(V(\psi))$, where
  $V(\psi)$ is the set of variables mentioned in $\psi$, and
  $\tau_\psi(x)=\cset{A\in X_\psi : x\in A}$
  (Remark~\ref{rem:os-univ}). We then construct the carrier $X_\phi$
  of $C_\phi$ as a subset of $\Pow(V)$. For $A\subseteq
  V$, we let $\phi_A$ denote the conjunctive one-step formula
  given by $\Land_{x_i \in A} \phi_i$.  Then, $X_\+\phi$ is the
  smallest subset of $\fP(V)$ containing $r_\phi = \cset{x_0}$
  such that
    $X_{\phi_A}\subseteq X_\phi$, for each $A\in X_\phi$.
  We define a $T$-coalgebra structure $\xi_\phi$ on $X_\phi$ by
    $\xi_\phi(A)=T(i_A)t_{\phi_A}$,
  where $i_A$ is the inclusion $X_{\phi_A}\into X_\phi$.
  \begin{theorem}\label{thm:univ-gfp}
    If $\Lambda$ admits one-step wsi models, then for every
    shallow $L^\nu(\Lambda)$-sentence $\phi$, $(C_\phi,r_\phi)$ as
    constructed above is a wsi model.
  \end{theorem}
  \begin{proof*}{Proof (Sketch)}
    Let $\phi$ have the form $\nu(x_0;x_1,\ldots
    x_n).(\phi_0;\phi_1,\ldots,\phi_n)$, so $V = \cset{x_0,\dots
      x_n}$.  We have to show that (i) $r_\phi \models_{C_\phi} \phi$
    and (ii) that if $d \models_D \phi$, then $r_\phi S d$ for some
    simulation $S : C_\phi \to D$ (Remark~\ref{rem:canonical-alt}).

    \emph{(i):} By coinduction -- taking $\+V(x_i) = \cset{A \in
      X_{\phi} : x_i \in A}$, one shows that $\+V(x_i)
    \subseteq \bbrack{\phi_i}_{C_\phi,\+V}$ for all $x_i \in V$. The
    gfp property of $\phi$ then implies
    $\+V(x_0)\subseteq\bbrack{\phi}$, and clearly $r_\phi\in\+V(x_0)$.


    \emph{(ii):} For $i = 0,\dots n$, let $\phi^{(i)}$ denote the
    formula obtained by projecting the $i$-th component of $\phi$:
 $$\phi^{(i)}\!=\!\nu(x_i;x_0\dots x_{i-1},x_{i+1}\dots x_n).(\phi_i;\phi_0\dots \phi_{i-1},\phi_{i+1}\dots\!\,),$$
 so in particular $\phi^{(0)} = \phi$.
Assume $d \models_D \phi$ for some coalgebra $D = (Y,\zeta)$.
Define a relation $S\subseteq X_\phi \times Y$ by
 \begin{equation*}
   ASy\iff y \models_D\Land_{x_i \in A}\phi^{(i)}.
 \end{equation*}
 Then clearly $r_\phi Sd$, for by definition $r_\phi=\cset{x_0}$ and
 $\phi^{(0)} = \phi$. One can show that $S$ is a $\Lambda$-simulation.
  \end{proof*}
  \noindent Clearly, all conjunctive logics listed as having one-step
  wsi models in the examples of Sec.~\ref{sec:wsi-models}
  have wsi models when extended with greatest fixpoints, in particular
  remain convex.
  Of course, the wsi models constructed above may be exponentially large,
  even when $\Lambda$ admits linear one-step wsi models.
  However, under $k$-boundedness, elements of $X_\phi\subseteq\Pow(\Delta)$ have
  at most $k$ elements, leading to our main criterion for smallness of
  wsi models under greatest fixpoints:
  \begin{theorem}\label{thm:univ-nu}
    If $\Lambda$ admits $k$-bounded one-step wsi models for
    some $k$, then conjunctive $\Lambda$-$\nu$-formulas have polymomial-size
    wsi models.
  \end{theorem}

\noindent
 By Theorem~\ref{thm:univ-nu} and the description of one-step
 wsi models in Section~\ref{sec:wsi-models}, and using
 abstract wsi models as in Section~\ref{sec:tractability},
 we regain the known result that subsumption checking over
 classical TBoxes with gfp semantics in \EL is in
 $P$~\cite{BaaderIJCAI03a}, and in fact can extend it to allow a
 bounded number of universal restrictions, always in conjunction
 with $\Diamond\top$. As new results, we obtain:

\begin{corollary}
 Subsumption checking for conjunctive Game Logic is in $P$.
\end{corollary}
\begin{cor}
  Subsumption checking for the conjunctive alternating-time
  $\mu$-calculus (AMC) without $[\emptyset]$ and $\tup{N}$ is in $P$.
\end{cor}

  \begin{remark}
    There is one case where we do obtain polynomial-size wsi
    models without $k$-boundedness, namely $\Lambda=\{\Box\}$ over
    Kripke frames -- here, one-step wsi models have only one
    state, so that wsi models for fixpoint formulas are
    lassos, i.e.\ chains of states ending in a loop. For smallness,
    one still needs to impose additional restrictions on shallow
    fixpoints $\nu(y;y_1,\dots,y_n).(\phi;\phi_1,\dots,\phi_n)$, e.g.\
    that $y_i$ always appears in $\phi_i$, or that the fixpoint is
    acyclic, i.e.\ not actually recursive. This example does not
    extend to the multi-modal case since the property of one-step
    wsi models being singletons is not stable under taking
    disjoint sums (Remark~\ref{rem:univ-fusion}). Indeed, the
    multimodal version is $FL$, and reasoning over even
    the most restrictive (i.e.\ acyclic) TBoxes in \FL is
    known to be coNP-hard~\cite{Nebel90}.
  \end{remark}

\section{Conclusions}

\noindent Representability of formulas by models in the sense that
simulation of the model is equivalent to satisfaction of the formula
is a highly useful phenomenon in conjunctive fragments of modal fixpoint
logics. It implies, for instance, convexity of the formula (and is equivalent
to it in the one-step case) and
under a polynomial size bound on the model, tractability of reasoning.
We have studied the question of
existence of such \emph{weakly simulation-initial (wsi) models}, in
the framework of coalgebraic logic; in particular, we have proved a
reduction of the problem to a local (\emph{one-step}) version.
We were able to derive a criterion for tractability from the shape
of the tableau rules that enabled us to establish tractability
in a number of key examples:
\begin{itemize}
\item we have recovered known tractability results for the description
  logics \EL (over classical TBoxes with gfp semantics) and \FL
  (without a TBox), and shown that reasoning over classical TBoxes
  with gfp semantics in \EL (equivalently in the fragment of the
  multi-modal $\mu$-calculus defined by restricting to conjunction,
  diamonds, and greatest fixed points) remains tractable when we allow
  a bounded number of universal restrictions (i.e.\ boxes);
\item we established tractability of conjunctive monotone
  logic with greatest fixed points over serial models, which subsumes
  corresponding fragments of game logic~\cite{Parikh83};
\item we have shown tractability of the conjunctive fragment (which
  has greatest but not least fixed points) of the alternating-time
  $\mu$-calculus AMC~\cite{AlurEA02}; this fragment still includes the
  game-based versions of $EG$ and $AG$ found in ATL.
\end{itemize}
\noindent Outside the large body of work on \EL, there has been only a
limited amount of research on wsi models for conjunctive
logics.  Notable examples are the work on the relationship between
relational modal logics and modal transition
systems~\cite{BoudolLarsen92,AcetoEA11} where formulas in certain
variants of positive Hennessy-Milner logic are shown to have wsi
models iff they are convex. (\emph{prime} in the cited works).  We
exhibited a similar equivalence at the level of conjunctive
coalgebraic \emph{logics}; we leave a generalization of the
equivalence for individual \emph{formulas} as future work.
There is some work on sub-Boolean fragments of temporal
logics, which however focuses on satisfiability rather than
subsumption (e.g.~\cite{MeierEA09}).

Further points for future investigation include the use of
wsi models to calculate so-called \emph{least common
  subsumers}~\cite{BaaderKuesters99}, as well as covering
\emph{general} TBoxes (i.e.\ finite sets of arbitrary inclusion
axioms), which is known to remain tractable in the case of
\EL~\cite{Brandt04}.

\paragraph*{Acknowledgments} The authors wish to thank Carsten Lutz
for useful discussions, and Erwin R.\ Catesbeiana for unsolicited
remarks regarding the absence of unsatisfiable formulas in \EL.

\bibliographystyle{aiml14}
\bibliography{aiml14}

\newpage\appendix

\section{Appendix: Omitted Proof Details}

\subsection*{Details on Notation in the One-step Logic}

One-step $\Lambda$-formulas are parametrized over a set $V$ of
propositional variables, which for purposes of the current work can
w.l.o.g.\ be assumed to be always finite. Given any set $Z$, we denote
by $\Conj(Z)$ the set of finite conjunctions over $Z$, by $\Pos(Z)$
the set of positive propositional formulas over $Z$, and by
$\Lambda(Z)$ the set $\{\M z : M\in\Lambda,z\in Z\}$. Recall that
positive formulas over $V$ have disjunctive normal forms consisting of
conjunctions over $V$. A \emph{positive $\Lambda$-formula over $V$} is
an element of $\Pos(\Lambda(\Pos(V)))$, and a \emph{conjunctive
  $\Lambda$-formula over $V$} an element of $\Conj(\Lambda(V))$.

A \emph{one-step model} $(X,\tau,t)$ over $V$ consists of a set $X$, a
valuation $\tau:V\to\Pow(X)$, and $t\in TX$. For $\rho\in\Pos(V)$
(including $\rho\in\Conj(V)$), we denote the extension of $\rho$ under
$\tau$ in the Boolean algebra $\Pow(X)$ by $\Sem{\rho}_\tau$. For
$\M\in\Lambda$, $\rho\in\Pos(V)$, we put
$\Sem{\M\rho}_\tau=\Sem{\M}(\Sem{\rho}_\tau)\subseteq TX$, and extend
this to define $\Sem{\phi}_\tau\subseteq TX$ for positive
$\Lambda$-formulas $\phi$ using the Boolean algebra structure of
$\Pow(TX)$. We write $(X,\tau,t)\models\phi$ if $t\in\Sem{\phi}_\tau$.
For positive $\Lambda$-formulas $\phi,\psi\in\Pos(\Lambda(\Pos(V)))$,
we write $\phi\ossub\psi$ if $(X,\tau,t)\models\psi$ whenever
$(X,\tau,t)\models\phi$.

In this notation, a \emph{monotone one-step tableau rule}
$\Gamma_0/\Gamma_1\dots\Gamma_n$ over $V$ consists of a
\emph{conclusion} $\Gamma_0\in\Conj(\Lambda(V))$, mentioning every
variable in $V$ at most once, and \emph{premises}
$\Gamma_1\dots\Gamma_n\in\Conj(V)$, all read as sets of positive
literals; we require that every variable occurring in one of the
conclusions occurs also in the premise. The reading of such a rule is
the usual one for tableau rules -- to show that a set of formulas is
satisfiable, show that all matching tableau rules have at least one
satisfiable conclusion. Soundness and completeness can be reduced to
the one-step level: a rule $\Gamma_0/\Gamma_1\dots\Gamma_n$ over $V$
is \emph{one-step sound} if for every $\tau:V\to\Pow(X)$,
$\Sem{\Gamma_0}_\tau\neq\emptyset$ implies that
$\Sem{\Gamma_i}_\tau\neq\emptyset$ for some $i\in\{1,\dots,n\}$; and a
set $\Rules$ of monotone one-step tableau rules is \emph{one-step
  tableau complete} if for all $\phi\in\Conj(\Lambda(W))$ (again read
as a set of literals) and all $\tau:W\to\Pow(X)$, we have
$\Sem{\phi}_\tau\neq\emptyset$ whenever for all rules
$\Gamma_0/\Gamma_1\dots\Gamma_n$ over $V$ in $\Rules$ and all
 substitutions $\sigma:V\to W$ such that
$\Gamma_0\sigma\subseteq\phi$, there exists $i\in\{1,\dots,n\}$ such
that $\Sem{\Gamma_i\sigma}_\tau\neq\emptyset$.

\subsection*{Proof of Theorem~\ref{thm:univ}}

Induction on the modal depth of $\phi$. We have $\phi = \Land_{i \in
  I}\M_i\chi_i$ for a finite (possibly empty) index set $I$.  Take
$\Var_\phi = \cset{a_{\chi_i} : i \in I}$ and decompose $\phi$ as
$\phi=\phi^*\rho$ into a one-step formula $\phi^* = \Land_{i\in I}\M_i
a_{\chi_i}$ and a substitution $\rho(a_{\chi_i}) = \chi_i$.  Let
$(X,\tau,t)$ be wsi for $\phi^*$. By
induction, we have, for each $x \in X$, a wsi model $(C_x,x)$
with $C_x=(Y_x,\xi_x)$ for $\Land_{p \in\breve\tau(x)}\rho(p)$ with
root $x$. We assume w.l.o.g.\ that the $Y_x$ are pairwise
disjoint. Pick a fresh $x_\phi$, and put
$Y=\{x_\phi\}\cup\bigcup_{x\in X}Y_x$. Let $i_X:X\into Y$ and
$i_{Y_x}:Y_x\into Y$ denote the respective subset inclusions. Define a
$T$-coalgebra $C_\phi=(Y,\xi)$ by $\xi(x_\phi)=Ti_X(t)$ and
$\xi(y)=Ti_{Y_x}(\xi_x(y))$ for $y\in Y_x$.

We claim that $(C_\phi,x_\phi)$ is wsi for
$\phi$. One shows by induction that $x_\phi\models_C\phi$ (this is
essentially as in~\cite{SchroderPattinson08}). It remains to show that
given a coalgebra $D = (Z,\zeta)$ and $z_0\in Z$ such that
$z_0\models_D\phi$, $z_0$ $\Lambda$-simulates $x_\phi$. Define a
one-step model $(Z,\theta,\zeta(z_0))$ by putting
$\theta(a)=\Sem{\rho(a)}_D$. Then $\zeta(z_0)\models_\theta\phi^*$, so
we have a one-step $\Lambda$-simulation $S$ between $t$ and
$\zeta(z_0)$ such that $S[\tau(a)]\subseteq\theta(a)$ for all
$a$. This implies that whenever $xSz$ then $z\models_D\Land_{a\in
  \Var_\phi}\rho(a)$, so that there exists a $\Lambda$-simulation
$S_{xz}$ between $(C_x,x)$ and $(D,z)$. Then
$R=\{(x_\phi,z_0)\}\cup\bigcup_{xSz}S_{xz}$ is a $\Lambda$-simulation
between $(C_\phi,x_\phi)$ and $(D,z_0)$. \qed

\subsection*{Proof of Lemma~\ref{lem:os-mat}}

The proof that one-step wsi models are one-step
materializations is as for the full logic (Lemma~\ref{lem:min}). To
show the converse implication, let $(X,\tau,t)$ be a one-step
materialization of a conjunctive one-step $\Lambda$-formula $\phi$
over $V$, w.l.o.g.\ with $V$ finite. Let $(Y,\theta,s)\models\phi$,
let $t\models \M A$, and let $xSy$ iff
$\breve{\tau}(x)\subseteq\breve{\theta}(y)$. We have to show that
$s\models\M S[A]$. Define $\rho\in\Pos(V)$ as $\rho=\Lor_{x\in A}
\Land \breve{\tau}(x)$ (effectively a finite disjunction since $V$ is
finite). Then $A\subseteq\Sem{\rho}_\tau$. Thus,
$(X,\tau,t)\models\M\rho$, and hence $\phi\ossub\M\rho$, so that
$(Y,\theta,s)\models\M\rho$. We are done once we show that
$\Sem{\rho}_\theta\subseteq S[A]$. So let $y\in\Sem{\rho}_\theta$. By
definition of $\rho$, there exists $x\in A$ such that
$y\in\Sem{\Land\breve{\tau}(x)}_\theta$. Then
$\breve{\tau}(x)\subseteq\breve{\theta}(y)$, so $y\in S[A]$. \qed

\subsection*{Proof of Lemma~\ref{lem:os-convex}}

`Only if' is clear; we prove `if'. Thus, let $\phi$ be a strongly
convex positive one-step $\Lambda$-formula. Then $\phi$ is equivalent
to one of the conjunctive clauses in its DNF, so we can assume
$\phi\in\Conj(\Lambda(\Pos(V)))$; we understand $\phi$ as a subset of
$\Lambda(\Pos(V))$. To show that $\phi$ is materializable,
it suffices to show that the set $\phi\cup\tilde\phi$, where
\begin{equation*}
  \tilde\phi=\{\neg\M\rho : \M\rho\in\Lambda(\Pos(V)),\phi\not\sqsubseteq\M\rho\}
\end{equation*}
(with the expected semantics of negation) is one-step satisfiable,
i.e.\ satisfied in some one-step model. Assume the contrary; that is,
$\phi\sqsubseteq\Lor\{\M\rho : \neg\M\rho\in\tilde\phi\}$. By strong
convexity, we then have $\phi\sqsubseteq\M\rho$ for some
$\neg\M\rho\in\tilde\phi$, contradiction.\qed

\subsection*{Proof of Theorem~\ref{thm:convex-rules}}

To begin, we state the result in greater generality, as indicated in
brackets in the main text. We recall the relevant strengthened notion
of one-step completeness~\cite{SchroderPattinson09b}:
\begin{defn}\rm
  The set $\Rules$ of rules is \emph{strongly one-step tableau
    complete over finite sets} if for all $\phi\subseteq\Lambda(W)$
  with $W$ finite (but $\phi$ possibly infinite, read as an infinite
  conjunction) and all $\tau:W\to\Pow(X)$, we have
  $\Sem{\phi}_\tau\neq\emptyset$ (with $\Sem{\phi}_\tau=\bigcap_{\M
    a\in\phi}\Sem{\M a}_\tau$) whenever for all rules
  $\Gamma_0/\Gamma_1\dots\Gamma_n$ over $V$ in $\Rules$ and all
  substitutions $\sigma:V\to W$ such that
  $\Gamma_0\sigma\subseteq\phi$, there exists $i\in\{1,\dots,n\}$ such
  that $\Sem{\Gamma_i\sigma}_\tau\neq\emptyset$.
\end{defn}
Clearly if $\Lambda$ is finite then $\Rules$ is strongly one-step
tableau complete over finite sets iff it is one-step tableau complete,
so the following version of the theorem does generalize the one stated
in the main text. Requiring $W$ to be finite is equivalent to
restricting $X$ to be finite (if $X$ is finite, then $\phi$ can be
equivalently transformed, over $(X,\tau)$, into a set of formulas
mentioning only finitely many variables; conversely, if $W$ is finite,
then we can quotient $X$ by the kernel of $\breve{\tau}$ and obtain a
finite set); hence the term `over finite sets'.

\subsection*{Proof of Theorem~\ref{thm:convex-rules}}
\newcommand{\Neg}{\mathsf{Neg}}
  As in the proof of Lemma~\ref{lem:os-convex}, put
  \begin{equation*}
  \tilde\phi=\{\bar\M\neg\rho : \M\rho\in\Pos(V),\phi\not\sqsubseteq\M\rho\}.
  \end{equation*}
  It suffices to show that $\phi\cup\tilde\phi$ is one-step
  satisfiable over $(X,\tau)$, i.e.\ that there exists $t\in TX$ such
  that $(X,\tau,t)\models\phi\cup\tilde\phi$. Let $\Gamma/\Delta$ be a
  (by assumption, definite) rule over $W$ in $\Rules$, and let
  $\sigma$ be a substitution such that
  $\Gamma\sigma\subseteq\phi\cup\tilde\phi$. By strong one-step
  tableau completeness over finite sets, it suffices to show that
  $\Sem{\Delta\sigma}_\tau\neq\emptyset$. Note that we have a
  decomposition $\Gamma=\Gamma_0,\Gamma_1$ such that
  $\Gamma_0\sigma\subseteq\phi$ and
  $\Gamma_1\sigma\subseteq\tilde\phi$. This induces a disjoint
  decomposition $W=W_0\cup W_1$, and then a disjoint decomposition
  $\Delta=\Delta_0,\Delta_1$ with $\Delta_0=\Delta\cap W_0$,
  $\Delta_1=\Delta\cap W_1$. By preservation of $\Lambda$-convexity,
  we then have a rule of the form
  $\Gamma_0,\Gamma_1'/\Delta_0,\Delta_1'$ with
  $\Gamma_1'\subseteq\Gamma_1$, $\Delta_1'\subseteq\Delta_1$, and
  $|\Gamma_1'|,|\Delta_1'|\le 1$ (more precisely, this is by
  preservation of $\Lambda$-convexity if $\Delta_1\neq\emptyset$, and
  trivial otherwise). By construction of $(X,\tau)$, we thus have
  $x_0\in X$ with $\breve{\tau}(x_0)=\Delta_0\sigma$. Now assume
  $\Sem{\Delta\sigma}_\tau=\emptyset$. Then in particular
  $x_0\notin\Sem{\Delta\sigma}_\tau$. Since $x_0$ makes just enough
  propositional variables true to satisfy $\Delta_0\sigma$, and
  $\Delta_1\sigma$ is the negation of a positive formula, this implies
  that $\Delta\sigma=\Delta_0\sigma,\Delta_1\sigma$ is actually
  unsatisfiable. Since $\Delta_0\sigma$ is a conjunction of
  propositional variables and hence convex in positive propositional
  logic, it follows that already $\Delta_0\sigma,\sigma(a)$ is unsatisfiable
  for some $a\in\Delta_1$. Since $\Rules$ preserves
  $\Lambda$-convexity, this implies that
  $\Gamma_0\sigma,\bar\M\sigma(a)$ is unsatisfiable for the unique
  $\M\in\Lambda$ with $\bar\M a\in\Gamma_1$. Since
  $\Gamma_0\sigma\subseteq\phi$, it follows that
  $\phi\sqsubseteq\M\neg\sigma(a)$, in contradiction to
  $\bar\M\neg\rho\in\tilde\phi$.

\subsection*{Remark on Simultaneous Fixpoints}

\begin{remark}\label{rem:simultaneous-fp}
 Note that $\nu(x_0;x_1,\dots,x_n).(\phi_0;\phi_1\dots\phi_n)$ is a
 simultaneous fixpoint in which only the extension of $x_0$ is projected.
 More precisely, for $i = 0,\dots n$, let $\phi^{(i)}$ denote the formula
 obtained by projecting the $i$-th component of $\phi$:
 $$\phi^{(i)}\!=\!\nu(x_i;x_0\dots x_{i-1},x_{i+1}\dots x_n).(\phi_i;\phi_0\dots \phi_{i-1},\phi_{i+1}\dots\!\,)$$
 so in particular, $\phi^{(0)} = \phi$. It then follows from the
 definition of the semantics of $\nu$ and monotonocity of the fixpoint
 arguments that for $i = 0,\dots n$,
\begin{equation*}\label{eq:fp-unfold}
 \bbrack{\phi_i^{(i)}}_{C,\+V}
   = \bbrack{\phi_i}_{C,\+V[x_0 \mapsto \phi^{(0)}\dots x_n \mapsto \phi^{(n)}]}
\end{equation*}
which incidentally gives us a rule for unfolding fixpoints.
\end{remark}

\subsection*{Proof of Theorem~\ref{thm:univ-gfp}}
 Recall that $\phi$ is assumed to be a shallow formula
 of the form $\nu(x_0;x_1,\ldots x_n).(\phi_0;\phi_1,\ldots,\phi_n)$,
 so $V = \cset{x_0,\dots x_n}$.
 It is enough to show that (i) $r_\phi \models_{C_\phi} \phi$
 and (ii) that if $d \models_D \phi$, then $r_\phi S d$
 for some simulation $S : C_\phi \to D$ (Remark~\ref{rem:canonical-alt}).

 \emph{(i):} We will show a slightly stronger result, namely that for
 $\+V(x_i) = \cset{A \in X_{\phi} : x_i \in A}$, and for all $x_i \in V$
  it holds that $\+V(x_i) \subseteq \bbrack{\phi_i}_{C_\phi,\+V}$.

  So let $A \in \+V(x_i)$ and let $\M x_j$ be any of the conjuncts of $\phi_i$.
  We only need to verify that $\xi_\phi(A) \models \M_{X_\phi}\+V(x_j)$.
  Now, by definition $(X_{\phi_A},t_{\phi_A},\tau_{\phi_A}) \models \phi_A$
  and, in particular, $t_{\phi_A} \models \M\tau_{\phi_A}(x_j)$. Moreover,
  $\tau_{\phi_A}(x_j) = \+V(x_j) \cap X_{\phi_A}$ and by naturality of $\M$
  we get $\xi_\phi(A) = T(i_A)t_{\phi_A} \models \M\+V(x_j)$.

\emph{(ii):}
Assume $d \models_D \phi$ for some coalgebra $D = (Y,\zeta)$.
Define a relation $S\subseteq X_\phi \times Y$ by
 \begin{equation}\label{eq:sim-from-sat-gfp}
   ASy\iff y \models_D\Land_{x_i \in A}\phi^{(i)}
 \end{equation}
where $\phi^{(i)}$ is as in Remark~\ref{rem:simultaneous-fp}.
Then clearly $r_\phi Sd$, for by definition $r_\phi=\cset{x_0}$
and $\phi^{(0)} = \phi$.
We claim that $S$ is a $\Lambda$-simulation.
So let $ASy$, and let $\xi_\phi(A)\models\M\FB$ for some $\M\in\Lambda$
and $\FB\subseteq X_\phi$. We then have to show $\zeta(y)\models\M S[\FB]$.
Since $\xi_\phi(A)=Ti_A  (t_{\phi_A})$, we have
$t_{\phi_A}\models\M(\FB\cap X_{\phi_A})$ by naturality of predicate liftings.
 Since $ASy$, it follows from~\eqref{eq:sim-from-sat-gfp} and
 \eqref{eq:fp-unfold} that
$\zeta(y)\models_\theta\phi_A$ where $\theta(x_i)=\Sem{\phi^{(i)}}_D$ for all
 $x_i\in V$. But $(X_{\phi_A},\tau_{\phi_A},t_{\phi_A})$ is
 wsi for $\phi_A$, so it follows that $\zeta(y)\models\M
 R[\FB\cap X_{\phi_A}]$ where $BRz\iff \breve\tau_{\phi_A}(B)\subseteq\breve\theta(z)$
 (recall that by definition $\breve\tau_{\phi_A}(B)=B$).
 Now  $R[\FB\cap X_{\phi_A}]\subseteq S[\FB]$: when $B\in\FB\cap
 X_{\phi_A}$ and $BRz$, then $z \in \bbrack{\phi^{(i)}}_D$ for each
 $x_i \in B$ so that $BSz$.
 By monotonicity, we obtain  $\zeta(y)\models\M S[\FB]$ as required.
 \qed

\end{document}